# SAXSDOG: open software for real-time azimuthal integration of 2D scattering images

*Max Burian[1,a], Christian Meisenbichler[1,b], Denys Naumenko[1] and Heinz Amenitsch[1,*]*

[1] Institute of Inorganic Chemistry, Graz University of Technology, Stremayergasse 9/V, Graz, 8700, Austria.

[a] current address: Paul Scherrer Institute, Forschungsstrasse 111, Villigen PSI, 5232, Switzerland.

[b] current address: AVL LIST GmbH, Hans-List-Platz 1, Graz, 8020, Austria.

[*] Correspondence email: heinz.amenitsch@elettra.eu

**ABSTRACT**

In-situ small- and wide-angle scattering experiments at synchrotrons often result in massive amounts of data within seconds only. Especially during such beamtimes, processing of the acquired data online, so without mentionable delay, is key to obtain feedback on failure or success of the experiment. We thus developed SAXSDOG, a python based environment for real-time azimuthal integration of large-area scattering-images. The software is optimized for dedicated data-pipelines: once a scattering image is transferred from the detector onto the storage-unit, it is automatically integrated and pre-evaluated using integral parameters within milliseconds. The control and configuration of the underlying server-based processes is done via a graphical user interface SAXSLEASH, which visualizes the resulting 1D data together with integral classifiers in real time. SAXSDOG further includes a portable "take-home" version for users that runs on standalone computers, enabling its use for laboratory machines.

## 1. INTRODUCTION

Synchrotron radiation sources provide the high flux needed for *in-situ* scattering experiments with milli- and microsecond time resolution [1–3]. These experiments are fundamental to study physical, chemical and biological mechanisms occurring at the molecular nanometer-level, whereas especially small angle x-ray scattering (SAXS) is among the few techniques offering structural insight into these phenomena [4,5]. For standard (non-stroboscopic [6] and/or continuous-flow [7]) experiments, the best-achievable time resolution is limited by the readout time of the X-Ray detector, which is in the order of milliseconds [8–13]. However, even framerates of 100-1000 Hz cause massive amounts of data, considering that 2D detectors often consist of more than 1 Megapixel [14]. This raises the demands on the hard- and software used in the data-processing pipeline, which are the backbone of stable and efficient beamline operation.

In small angle x-ray scattering (SAXS) and powder diffraction, the experimentally recorded 2D scattering images have to be transformed into 1D scattering patterns by means of azimuthal integration [15–19]. When done manually, this operation can be time-consuming and hence cause unused dead-time of user-dedicated ring-operation. Several beamlines and synchrotrons have developed custom solutions to accelerate and/or automate this integration process, [20–25] resulting in impressive computational performance close to current hardware limits [26]. As, however, each beamline is unique in regard of (a) its user-base, (b) the performed experiments and hence (c) the demands on data-processing, we aimed at developing a framework that (i) operates automated on the beamline's data-backbone, (ii) is configurable via a user interface and (iii) processes data (close to) real-time. The side-benefit of this vision: the reactions/phenomena measured during the experiment can be monitored in online via data-classifiers [27,28], providing rapid feedback on the experimental-conditions without further, often time-consuming, data-evaluation.

In this work, we present SAXSDOG, an open source, python based program designed for fast, online azimuthal integration and pre-evaluation of 2D scattering images, which is currently in operation at the Austrian SAXS beamline at Elettra. SAXSDOG offers two modes of operation: 1) a "local server mode" that can be run on standalone computers and 2) a "remote server mode" incorporated in the data pipeline of our endstation under which it reaches its full potential. We explain the common subroutines of both operation modes and focus on the details for the implementation of SAXSDOG in a performance server network as found at common beamlines. We further show how the server process is controlled and configured via the Qt based graphical user interface SAXSLEASH that is also used to visualize the integrated data as well as the corresponding integral parameters. These merits will be emphasized by means of a given example, which will also demonstrate how SAXSDOG can help to get a first glimpse at the studied effects without extensive data-evaluation.

## 2. SPECIFICATIONS

The SAXSDOG package is written in the Python (v3.5) and has been developed in the Anaconda framework. During the development, special care has been taken to make the software, and the included dependencies, cross-platform compatible. A detailed list of all package versions for stable operation can be found in the detailed user manual that is distributed with the source code. The most fundamental packages are: (i) Qt4 (providing the GUI as well as the signaling protocol between threaded process), (ii) JSON schema (providing the data-standard on server and client), (iii) PyZMQ (providing the network-communication standard) as well as (iv) Pillow (providing the python image-processing library). The maintained version of SAXSDOG can be downloaded from GitHub at https://github.com/maxburian/SAXS_py3 and includes a user manual (local website) with step-by-step installation instructions and more detailed information on the source-code. We explicitly encourage users to participate in further code development via the GIT platform. The software can be used and is released free of charge under the GNU General Public License.

## 3. PRINCIPLES – THE SAXSDOG NETWORK

The functionality of the software package is based on a server-client based principle, summarized in the SAXSDOG network shown in Figure 1. In the following, the data- and control-flow of the pipeline are explained.

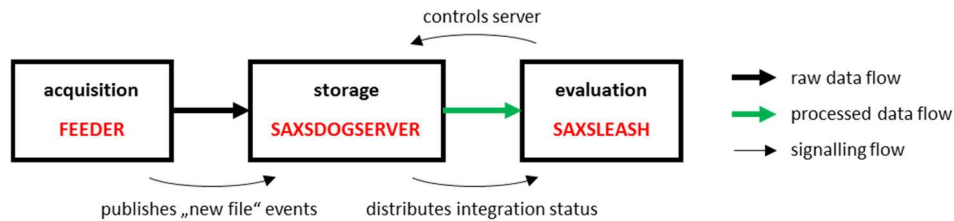

**Figure 1** Data- and signalling-flow of the SAXSDOG software package.

Once an image is acquired by the detector [here, Pilatus3 1M (Dectris, Switzerland)], it is automatically transferred from the temporary detector-storage [here, Pilatus Processing Unit (PPU)] to the beamline's (long-term) data-storage-server (see bold black arrow in Figure 1). This file-transfer script (here called *FEEDER*) includes a command that publishes a "new file" event over the network via ZMQ-message. On the data-storage-server, the *SAXSDOGSERVER* waits for such a "new file" event from the *FEEDER*, as it signals which image to integrate next. Once the integration is completed, the processed data together with additional data classifiers (see bold green arrow in Figure 1) is stored and distributed in real-time to the GUI on the client/user PC: the *SAXSLEASH*.

The core module of this processing pipeline is the *SAXSDOGSERVER*, a daemon-process running on the beamline's data-storage-server. The *SAXSDOGSERVER* is controlled via the *SAXSLEASH,* which sets the integration parameters and (de-)activates the processing queue. The underlying network is defined by the *$home/.saxsdognetwork* file (can be called via the *saxsnetconf* command), specifying the IP addresses of *FEEDER* and *SAXSDOGSEVER* as well as an authentication secret (which de- and encrypts the sent and received message, respectively). The precise functionality of all three modules is explained in the following subsections.

### 3.1. FEEDER

The *SAXSDOGSEVER* subscribes to the *FEEDER*: a script distributing "new file" events over the network. Such an event consists of a "command" ("new file") and an "argument" ("path\to\image\file\on\storage\server"), packaged in a Python dictionary. Once the image has successfully been copied from the temporary- to the data-storage-server, the "new file" command is rendered [(i) adding the remote file-path and (ii) encoding it to JSON) and sent via ZMQ (we use port 5555). In our implementation of SAXSDOG at the Austrian SAXS beamline, we have customized the GRIMSEL service provided by Dectris (Switzerland) that is responsible for transferring acquired images from the temporary-storage [Pilatus Processing Unit (PPU)] to the beamline's data-storage-server. It is important to note that the "new file" command is only sent once the image has been written successfully, as otherwise the *SAXSDOGSERVER* accesses a non-existing or non-complete file on the storage-server. An example of such a *FEEDER* script is shown in Figure 2.

```python
# importing libraries
import zmq, json, shutil, os

# setting-up ZMQ socket on Port 5555
context = zmq.Context()
socket = context.socket(zmq.PUB)
socket.bind("tcp://*:5555")

# initializing "new file" message
pub_msg_obj={"command":"new file","argument":""}

...

# copying file from src to remotedest
shutil.copy(src,os.path.dirname(remotedest))
# adjust command
pub_msg_obj["argument"] = remotedest
# send message
socket.send(json.dumps(pub_msg_obj))
```

**Figure 2**   Example of the core-code necessary to run the *FEEDER* service.

**3.2. SAXSDOGSERVER**

The *SAXSDOGSERVER* is the core module of the SAXSDOG software package as it performs all computational data-processing steps. A graphical overview of the working-principle is given in Figure 3, which is explained in detail in the following subsection.

The *SAXSDOGSERVER* is designed to run as a background service on the processing node of the data-storage server. When started, the process is idle, waiting for a *SAXSLEASH* to connect. The connection is only possible if the authentication secret de- and encrypting the network communication is identical on both machines. If the connection is established, the *SAXSDOGSERVER* waits for the integration-calibration, which defines geometry, integration-mask, directory, etc… (see section 4 for details). Once *SAXSLEASH* sends the "new" queue command, the actual integration process, the "image-queue", is initialized.

The image-queue is a threaded process that synchronizes two main modules: (a) the picture-queue and (b) the worker-pool (see Figure 3). In regard of (a), the picture-queue collects all filenames of the images that need to be processed in a central register. The picture-queue is filled by either (i) the *FEEDER* (for freshly acquired images) or (ii) by a directory walker (recursively identifying all existing images in the chosen folder path). In regard of (b), the worker pool consists of a user-defined number of parallel "workers" that perform the actual image-processing. Each worker takes one image after the other from the picture-queue, integrates it and stores the processed data (*.chi file). In addition to the image integration, the workers also calculate image classifiers [e.g. integral parameters [27,28]] of the scattering data and stores them in the *SAXSDOGSERVER* temporary-memory. This data may be queried from the *SAXSLEASH* at any moment, such that the integration progress can be monitored and preliminary data-evaluation is possible on-line, so simultaneously to image acquisition. The image-queue stays active until the "abort" command is sent from *SAXSLEASH* or until the *SAXSDOGSERVER* is terminated.

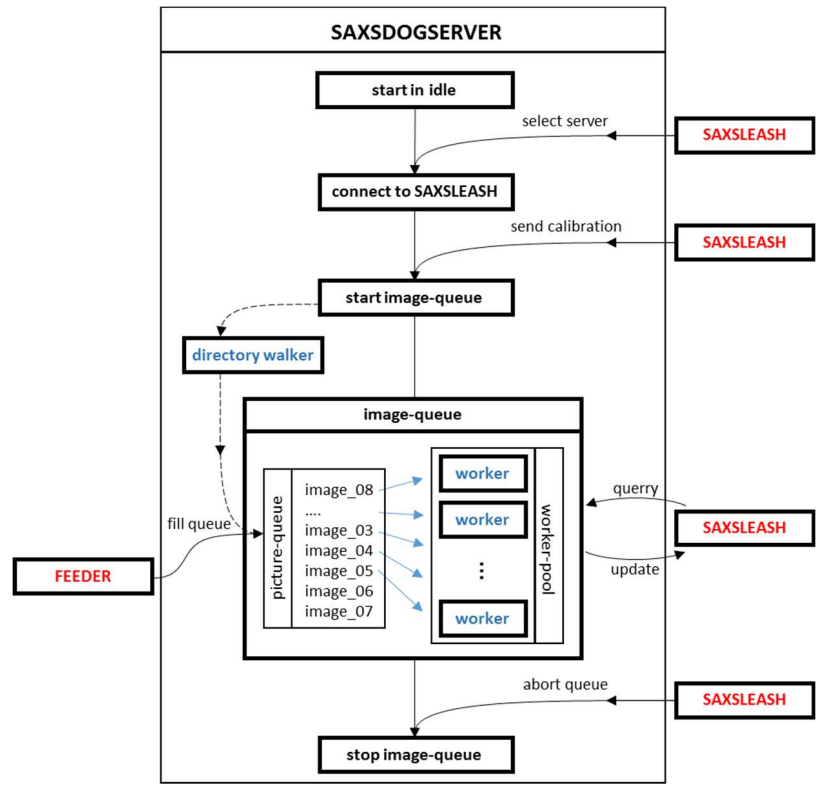

**Figure 3**   Internal workflow of the *SAXSDOGSERVER*. Segments in red refer to external processes (*FEEDER* and *SAXSLEASH*), segments in blue refer to internal processes (controlled by *SAXSDOGSERVER*). Lines with arrows indicate internal and external communication flow – dashed lines are optional procedures.

### 3.3. SAXSLEASH

The *SAXSLEASH* is a graphical user interface (GUI) that fulfills three main purposes: (i) setting up the integration-calibration, (ii) controlling the *SAXSDOGSERVER* and (iii) monitoring the integration status. In regard of (i), a calibration-editor allows to display and alter all required and optional integration-parameters (see Figure 4a) as well as select and display image-masking files (*.msk output from FIT2D). To simplify the input, we included converter tools such that the calibrated geometry-values from FIT2D [15] or NIKA2D [22] can be converted into the SAXSDOG format (see section 4 for details). In regard of (ii), the *SAXSDOGSERVER* may be controlled by (a) sending a new integration calibration, which starts a new image-queue, (b) forcing a reintegration of all existing image files (starts directory walker as shown in Figure 3) and (c) aborting and clearing the current image-queue. In regard of (iii), the current status of the integration on the *SAXSDOGSERVER* can be monitored in the plot- and history-tabs (see Figure 4b for history output on example data). From the histogram shown in the top-left, the integration progress as well as the integration speed are displayed. The other three panels show the image classifiers (here integral parameters) over the time the image was acquired (taken from image header), allowing a glimpse on e.g. reaction dynamics without further data-evaluation. Using the selection tool on the bottom of the window, the user can display the classifier-values of a single-dataset only, without having to reintegrate the entire image-queue.

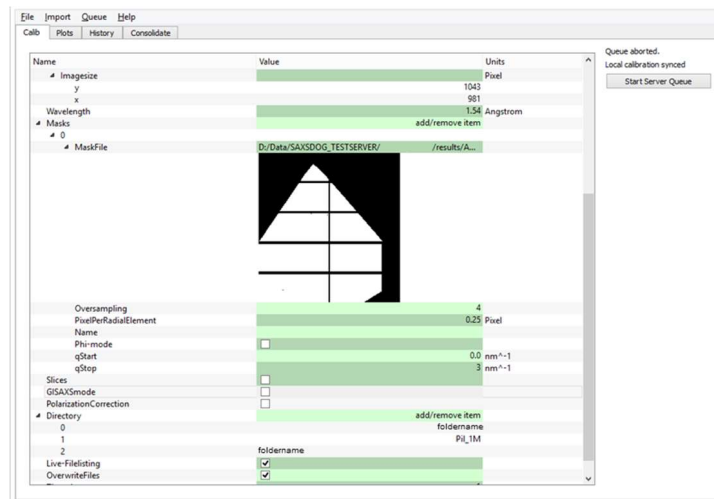

*(a)*

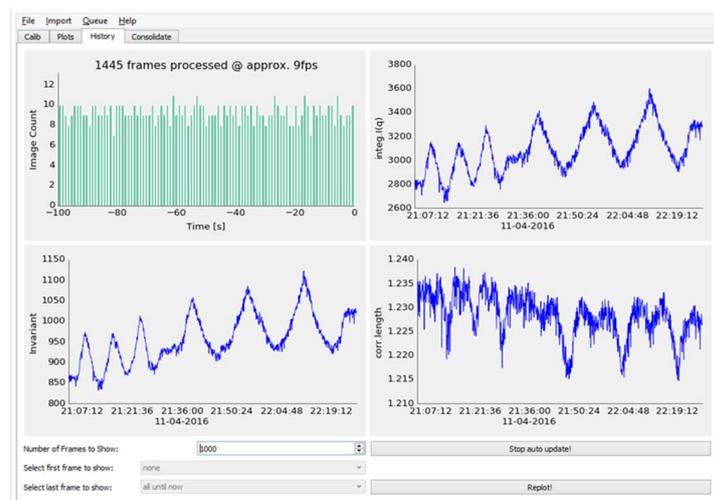

*(b)*

**Figure 4**  Screenshot of the *SAXSLEASH* GUI. (a) The calibration-tab allows the user to display and alter all required and optional integration-parameters. (b) The history-tab shows the current progress of the integration on the *SAXSDOGSERVER* [progress histogram (top-left)] and visualizes the calculated image-classifiers [integral-intensity (top-right), invariant (bottom-left) and correlation-length (bottom-right)].

## 3.4. LOCAL-SERVER-MODE

Instead of running all three SAXSDOG modules, so *FEEDER, SAXSDOGSERVER* and *SAXSLEASH* on separate machines, we implemented a "local-server-mode" that automatically emulates a working network on a single machine. The local-server can be selected as a starting option when *SAXSLEASH* is run. When selected, the user needs to specify a "working-directory", which acts as the root-directory of the hidden *SAXSDOGSERVER* process. The main difference to the dedicated implementation is that the local-server cannot be run together with the *FEEDER* (see Figure 3) such that only existing image-files can be integrated (only directory-walker fills image-queue – see Figure 3). However, this program option is ideal for beamline users to take home or for laboratory machines, where ease of use has higher priority than integration speed (see section 5 for performance metrics).

**Table 1** Mandatory parameters to be specified in the *calibration.*

| name | type | unit | description |
|---|---|---|---|
| *geometry* | object | | includes information on the experimental geometry |
| - *beamcenter* | array | pixel | position [vertical, horizontal] of the beamcenter on the detector |
| - *detector distance* | number | mm | sample to detector distance |
| - *image size* | array | pixel | dimensions [vertical, horizontal] of the sensor |
| - *pixel size* | array | μm | pixel size [vertical, horizontal] on the detector |
| - *tilt* | object | | includes information on the detector tilt |
| - - *tilt rotation* | number | degree | angle of the tilt direction |
| - - *tilt angle* | number | degree | angle between the primary beam and the normal of the detector |
| *masks* | array | object | list of masks to use for integration |
| - *path to file* | string | | path to mask file (supports *.msk files from FIT2D) |
| - *oversampling* | number | pixel | oversampling/anti-aliasing factor for radial integration |
| - *pix. p. rad. element* | number | pixel | width of each radial step in units of detector pixels |
| - *q-start* | number | nm$^{-1}$ | lower boundary for calculation of integral parameters |
| - *q-stop* | number | nm$^{-1}$ | upper boundary for calculation of integral parameters |
| *wavelength* | number | Å | wavelength of the X-Ray beam |
| *directory* | array | string | directory to take into account for processing images |
| *threads* | number | | number of parallel workers to use during image processing |

## 4. THE CALIBRATION

The *calibration* defines the parameters for image-processing on the *SAXSDOGSERER*. It includes all necessary information for azimuthal averaging, scattering-angle to scattering-vector conversion, image masking and processing options. A list of all required (mandatory) parameters can be found in Table 1 – please refer to the program manual for a description of optional parameters. The *calibration* is stored internally as a dictionary-type variable and is saved (as file) or communicated (sent via ZMQ from *SAXSLEASH* to *SAXSDOGSERVER*) according to the JSON structure. While the corresponding file for a specific experiment must hence be written in JSON-code, the *SAXSLEASH* provides a GUI for creating and editing such files without manual coding.

In order to better understand functionality of the integration parameters, the following subsections will explain the detector geometry, how the azimuthal integration is implemented and how horizontal- and vertical-slices can be used to evaluate GISAXS experiments.

### 4.1. THE GEOMETRY

For ease of operation, SAXSDOG uses the same detector geometry convention as FIT2D [15]. While parameters such as sample-to-detector-distance, beamcenter, image-size and pixel size are self-explanatory, special care has to be taken when working with tilted detectors. Here, we consider the trajectory of the normal to the detector plane, which is defined by two angles: (i) the *tilt rotation* $\varphi$ ("Rotating Angle of Tilting Plane" in FIT2D) and (ii) the *tilting angle* $\tau$ ("Angle of Detector Tilt in Plane" in FIT2D). A sketch of this geometry is shown in Figure 5a.

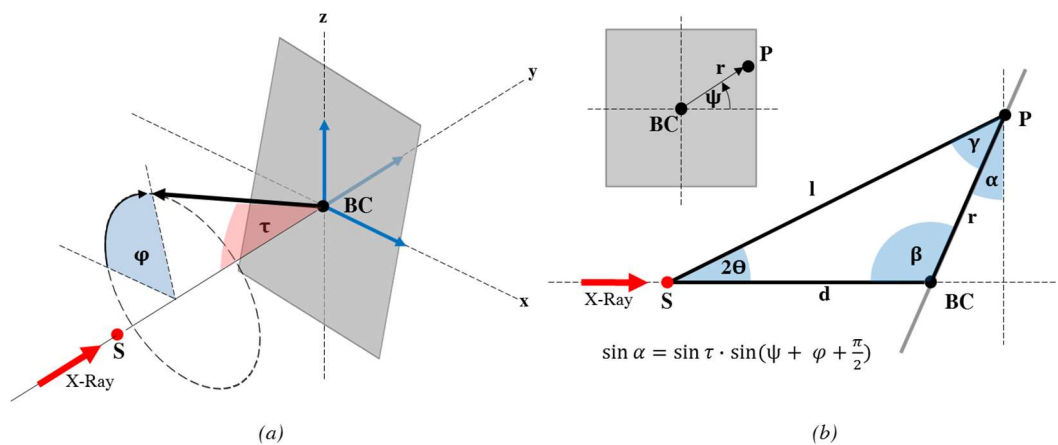

*(a)* *(b)*

**Figure 5** (a) Sketch of the geometry convention used in SAXSDOG. Here, the X-Ray primary beam (bold red arrow) incides on the sample (red dot) along the y-axis. The detector (grey plane) is tilted away from the primary-beam trajectory, as expressed by the sensor-plane normal (bold black arrow). This detector tilt can hence be defined in terms of two unique angles: (i) the *tilt rotation* $\varphi$ (blue arc) and (ii) the *tilting angle* $\tau$ (red arc). The distortion effects of these two tilting angles can, however, be summarized such that the 3D geometry can be reduced to a 2D geometric problem for each detector pixel in terms of $\alpha$ as shown in (b). Here, detector pixels are expressed in distance $r$ from the beamcenter (BC) and angle from vertical axis $\psi$ (see inset).

The 3D detector geometry is reduced to a 2D geometric problem by assigning two values to each detector pixel: (i) the distance $r$ between beamcenter (BC) and pixel (P) and (ii) the azimuthal angle $\psi$ between $r$ the vertical detector axis (see inset in Figure 5b). In this case, the detector *tilt rotation* $\varphi$, the detector *tilting angle* $\tau$ and the azimuthal angle $\psi$ can be summarized in terms of a single *distortion angle* $\alpha$ (see Figure 5b), which can be written as

$$\sin\alpha = \sin\tau \cdot \sin(\psi + \phi + \pi/2). \quad (1)$$

By knowing the distortion angle $\alpha$, the sample to detector distance $d$ and pixel-position $r$, the scattered light path $l$ can be obtained via

$$l = \sqrt{d^2 + r^2 - 2dr \cdot \cos(\pi/2 + \alpha)} \quad (2)$$

which is then used to determine the scattering angle $2\theta$ as well as the scattering vector magnitude $q$ according to

$$\cos 2\theta = \frac{l^2 - r^2 - d^2}{2ld} \text{ and } q = \frac{4\pi}{\lambda}\sin(\theta). \quad (3)$$

## 4.2. AZIMUTHAL INTEGRATION

The azimuthal integration in SAXSDOG is implemented in a matrix-vector multiplication scheme [similar to pyFai [29] but more simplified, as we do not perform angular regrouping]. Every image $\boldsymbol{p}$ [size:$(X, Z)$] is regarded as a 1D vector of pixels $\boldsymbol{p_i}$ [size:$(X \cdot Z, 1)$] such that the scattering intensity of each pixel is addressable by a single index $i$. In this scheme, the integration in a certain radial interval of the image can hence be seen as the weighted sum of all pixels: pixels within the radial element are weighted by 1 and pixels outside are weighted by 0. For a single radial element $j$, the mean intensity $I^j$ can be calculated by vector-vector multiplication (dot-product) of the weighting-vector $\boldsymbol{c^j}$ [size:$(X \cdot Z, 1)$] with the image-vector $\boldsymbol{p_i}$ via

$$I^j = \boldsymbol{c^j} \cdot \boldsymbol{p_i}. \quad (4)$$

As SAXSDOG intends to obtain all $N^j$ radial elements at once, this multiplication can be rewritten in the final matrix-vector dot-product form

$$\boldsymbol{I} = \boldsymbol{C} \cdot \boldsymbol{p_i} \quad (5)$$

where $\boldsymbol{C}$ [size:$(X \cdot Z, N^j)$] is the calibration-dependent weighting matrix and $\boldsymbol{I}$ [size:$(N^j, 1)$] is the azimuthally averaged scattering-intensity vector where the j[th] entry corresponds to the j[th] radial element (so $\boldsymbol{I_j} = I^j$). As most entries of the weighting matrix $\boldsymbol{C}$ are in fact zeros, we implement the computation in a sparse-matrix representation, which significantly speeds up computation and reduces the memory overhead.

The azimuthal integration also includes the calculation of the correct error band, here dominated by the random nature of scattering events that are best described by a Poisson distribution [30]. This Poisson error can be calculated for each radial element from (i) the mean intensity (obtained by azimuthal integration above) and (ii) the integration area of the radial slice. The integration area $A^j$ is obtained from the number of counted pixels within the j$^{th}$ radial element such that an area-vector $\mathbf{A}$ [size:$(N^j, 1)$] can be obtained in analogue to above by the matrix-vector dot-product of the weighting matrix $\mathbf{C}$ with a unit vector $\mathbb{1}$ [size:$(X \cdot Z, 1)$] by

$$\mathbf{A} = \mathbf{C} \cdot \mathbb{1}. \qquad (6)$$

The error vector $\mathbf{E}$ [size:$(N^j, 1)$] corresponding to the azimuthally averaged scattering-intensity $\mathbf{I}$ is then calculated for each image by elementwise multiplication (denoted as $*$) according to

$$\mathbf{E} = \sqrt{\mathbf{I} * \mathbf{A}^{-1}}. \qquad (7)$$

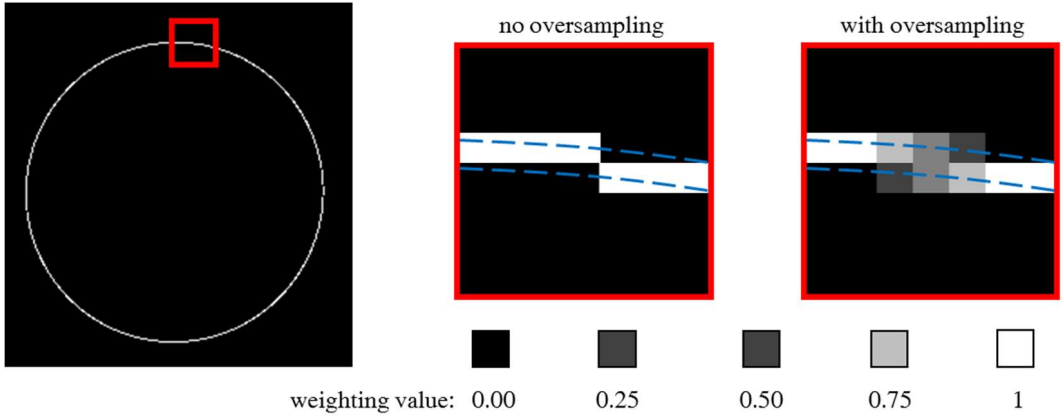

**Figure 6** Example of the problematic nature of discretization effects in azimuthal averaging of 2D detector images. The left shows all pixels in an example image that contribute to the azimuthal integration of a single radial element, whereas the red box highlights the region of interest. Here, integration within the geometrically defined radial segment (dotted blue lines) causes artefacts as in some cases it is not clear which pixels to count in- or outside of the integration area. Oversampling mitigates this issue, as non-binary weighting values allow that a single pixel can be considered for two radial segments but with corresponding weight.

The matrix-vector multiplication has been extended with an oversampling/anti-aliasing scheme that mitigates discretization effects in 2D image integration. An example of this oversampling scheme for a single radial element (here assuming no detector tilt) is shown in Figure 6. Considering a radial segment with width of a single pixel (see blue lines in Figure 6), one encounters two problems in azimuthal integration: (i) a single pixel might lie on the border of two segments and (ii) a radial segment might run through two pixels. By only choosing the nearest pixel for integration (see "no oversampling" in Figure 6), one may induce artefacts in the resulting curve, especially when only few pixels contribute (e.g.: close to the beamcenter in the small angle scattering region). Here, we use an algorithm similar to anti-aliasing in computer graphics [31] where we divide a much larger reference image (multiples of detector size) into the radial intervals and down sample the segments to the original image size. This yields a non-binary weighting matrix **C** for integration (see equation 5) that consists of intermediate weighting values between 0 and 1 such that the intensity is conserved after averaging.

### 4.3. GISAXS SLICES

For the rapid evaluation of experimental scattering data from grazing incidence SAXS (GISAXS) experiments, we included the option in SAXSDOG to calculate horizontal and vertical cuts of the detector image, so called *slices*. Such *slices* are defined as an array of *slice*-objects in the *calibration* – an overview of the necessary parameters can be found in Table 2. An example showing multiple slices within a single image is shown in Figure 7.

**Table 2** Parameters to be specified in the *calibration* to calculate the mean scattering intensity along a single slice.

| name | type | unit | description |
| --- | --- | --- | --- |
| *slices* | array | *slice* | array of *slice*-objects, as specified bellow |
| - *direction* | string |  | direction of the slice on the detector plane [x or y] |
| - *plane* | string |  | whether the slice-direction is in-plane with the scattering surface or perpendicular to it [InPlane or Vertical] |
| - *position* | number | pixel | pixel position where to place the slice (x-coordinate for y-*slice* and y-coordinate for x-*slice*) |
| - *margin* | number | pixel | number of pixels left and right from the position to include in the slice |
| - *mask reference* | number |  | pointer to the mask-object (see Table 1) to use for integration |

As known, grazing incidence experiments do not probe directly along the $q_Z$ direction in reciprocal space [32,33]. Moreover, in not-perfectly planar samples, the incidence angle (a critical parameter in performing the correct q-space conversion) is calculated from the specular-peak-position *after* measurements have been made and not at the moment when the integration calibration is defined. A single set of integration parameters that convert each image into 'true' reciprocal set parameters would hence induce a wrong q-scaling in the data-treatment pipeline and would make the *sliced*-data prone to misinterpretation. SAXSDOG therefore treats slices only in the detector coordinate system, so without Ewald-sphere distortion-correction (and neglecting the incidence angle of the X-Ray beam), such that it calculates the mean scattering intensity along the vertical- and horizontal-scattering components on the detector $q_V$ and $q_H$, respectively. It has to be noted that this assumption is only valid for small scattering angles or when samples are disordered in-plane and only partially aligned out-of-plane. For more ordered samples, we at this point refer to specialized programs for a correct and precise treatment of the q-space distortion for single images [34–36]. However, for the evaluation of GISAXS data during beamtimes, which is of peculiar interest for *in-situ* and *in-operando* experiments, the availability of *sliced* image-data through an automated data-pipeline is of high value by drastically facilitating the optimization of measurement conditions.

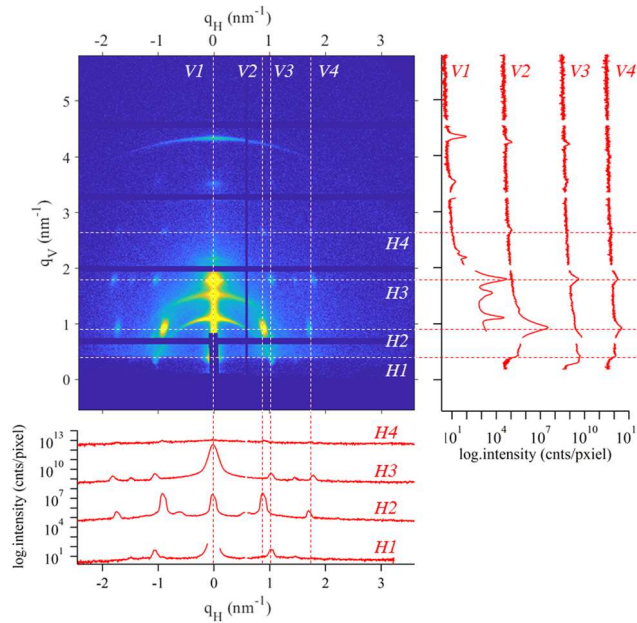

**Figure 7** Example of how multiple *slices* may be placed in detector images, here in a GISAXS pattern of nanostructured lipids on a mesoporous $SiO_2$ matrix. Horizontal *slices* (*direction: x* and *plane: InPlane*) and vertical slices (*direction: y* and *plane: Vertical*) are calculated within the detector system $q_H$ and $q_V$ and do not include the Ewald-Sphere distortion-correction. Dashed lines mark the *slice-position*, whereas for each slice a *margin* of 7 pixels (corresponding to a thickness of 2·7 + 1=15 pixels).

## 5. PERFORMANCE

We benchmark the performance of SAXDOG in terms of "frames per second" (fps) by measuring the total time required to integrate a set of images acquired with a Pilatus 1M (Dectris, Switzerland – 1Mega Pixel, uncompressed TIFF file format, 4 MB/file). The image acquisition rate was set at 200 Hz. The integration *calibration* included a dead-pixel mask only such that entire scattering-image is integrated and therefore the size of the sparse weighting matrix (see section 4.2) is kept at the maximum. The number of workers in the image-queue performing the image-processing (see Figure 3) is hence the main variable of this benchmarking test. For each server configuration three separate measurements (to estimate variability) consisting of 5.000 detector images each were made. Performance tests have been run on (i) the beamline server (dual socket, Intel Xeon E5-2650v4, 12-core @ 2.2 GHz, 24x10TB HDD Seagate ST10000NM0016) and (ii) a workstation laptop (Intel i7-4800MQ, 4-core @ 2.7 GHz, 1x500GB HDD Toshiba MQ01ACF050).

As seen in Figure 8 (red markers – "online"), the best performance (66.3 ± 4.9 fps) was achieved when using 64 workers, whereas with already 8-16 workers integration speeds of approx. 50 fps were observed. Peak processing-rates corresponds to a memory-access speed of ~280 MB/s, which is above the single-hard-drive hardware limit declared by the manufacturer of 250MB/s (enabled by the RAID data-storage system, such that image files are unconsciously read from different hard-drives). An increase of workers above 64 results in a drop in performance and is hence not recommended. The same performance measurements have been performed in the "offline" mode on the storage server, so when images have been acquired previously such that the image-queue is not filled by the *FEEDER* but by the directory-walker (see Figure 3). These measurements (see black markers in Figure 8) showed now mentionable performance-difference compared to the "online" integration mode. Data-acquisition by the detector as well as transfer onto the beamline server prior to integration hence do not seem to affect the integration performance. A long-term test, processing approx. 300.000 images (corresponding to 1.2 TB of data), showed that integration speeds of approx. 60 fps (using 40 workers – corresponding to 240 MB/s memory-access speed) can be maintained for more than 1.3 hours. When processing images on a normal workstation PC in "local-server" mode (see subsection 3.4) the overall integration speed is significantly lower (see blue markers in Figure 8): we achieve approx. 10 fps when using 4-16 image-queue workers. The memory-access rate of approx. 40 MB/s is half of the HDDs hardware limit, suggesting that either CPU or RAM performance are limiting the overall integration speed.

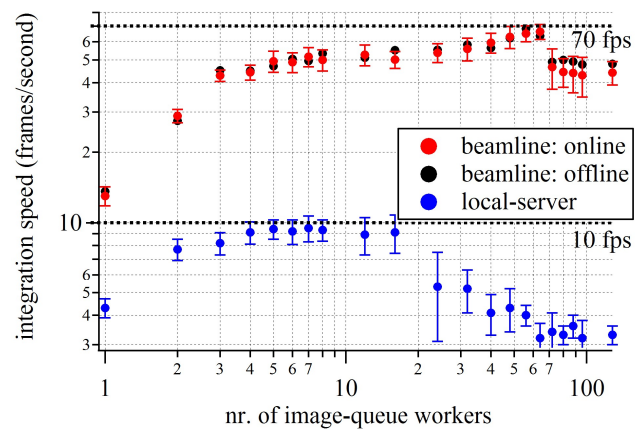

**Figure 8**  Results of the benchmarking tests of the SAXSDOG software on the Austrian SAXS beamline at the Elettra Sincrotrone storage ring. The integration speeds were determined from the time necessary to integrate 5.000 images from a Pilatus 1M detector (4 MB/image). In "online" mode, the image-queue is filled by the *FEEDER* and in "offline" mode, the image-queue is filled by the directory-walker (see Figure 3).

Overall, the performance of the SAXSDOG pipeline implemented at the Austrian SAXS beamline of the Elettra storage ring is sufficient to process scattering images "online", so within seconds after acquisition. The sustainable integration speed of approx. 60 fps (average processing time of 17 ms per image) ranks comparable to other software-packages [26] for azimuthal integration, especially considering that this rate (i) can be maintained over hours, (ii) already includes data transfer from the detector to the long-term data-storage system and (iii) is just barely below the hardware limit of the hard-drives (240 MB/s compared to manufacturer specified limit of 250 MB/s). For small datasets from e.g. lab-machines, processing-rates in the local-server mode are more than sufficient for azimuthal integration of SAXS data, making SAXSDOG and attractive choice also for not beamline related use.

## 6. CONCLUSION

In summary, we have designed, developed and implemented SAXSDOG: a software-package for fast, online integration of 2D scattering images. We show how the SAXSDOG-suite is useful for two separate operation schemes: 1) the "local server mode" that can be run on standalone computers and 2) the "remote server mode" as used in the data pipeline of the Austrian SAXS beamline at the Elettra synchrotron. By optimizing the program for online (real-time) integration during *in-situ* experiments, we reach peak integration performance at current hardware limits. Particular focus has been set on allowing operation via a graphical user interface, which sets all integration parameters, controls all ongoing server processes and shows the current integration progress, including image-classifiers for preliminary data-evaluation. The software (open-source code) can be used and is released free of charge under the GNU General Public License. We strongly encourage participation in further code development.

## ACKNOWLEDGEMENTS

The authors thank the IT group from the Elettra storage ring for aiding in implementation of the software on the server network, particularly: M.Biancho, R.Borghes and A.Curri.